\documentclass[aps,twocolumn,superscriptaddress]{revtex4}
\usepackage{amsmath}
\usepackage{graphicx}
\begin{document}

\title{The Fano resonance for Anderson impurity systems}

\author{H. G. Luo}
\affiliation{Institute of Theoretical Physics, Chinese Academy of
Sciences, P. O. Box 2735, Beijing 100080, China}
\author{T. Xiang}
\affiliation{Institute of Theoretical Physics, Chinese Academy of
Sciences, P. O. Box 2735, Beijing 100080, China} \affiliation{The
Interdisciplinary Center of Theoretical Studies, Chinese Academy
of Sciences, P. O. Box 2735, Beijing 100080, China}
\affiliation{Center for Advanced study, Tsinghua University,
Beijing 100084, China}
\author{X. Q. Wang}
\affiliation{Institute of Theoretical Physics, Chinese Academy of
Sciences, P. O. Box 2735, Beijing 100080, China} \affiliation{The
Interdisciplinary Center of Theoretical Studies, Chinese Academy
of Sciences, P. O. Box 2735, Beijing 100080, China}
\affiliation{Department of Physics, Renmin University of China,
Beijing 100872, China }
\author{Z. B. Su}
\affiliation{Institute of Theoretical Physics, Chinese Academy of
Sciences, P. O. Box 2735, Beijing 100080, China} \affiliation{The
Interdisciplinary Center of Theoretical Studies, Chinese Academy
of Sciences, P. O. Box 2735, Beijing 100080, China}
\affiliation{Center for Advanced study, Tsinghua University,
Beijing 100084, China}
\author{L. Yu}
\affiliation{Institute of Theoretical Physics, Chinese Academy of
Sciences, P. O. Box 2735, Beijing 100080, China} \affiliation{The
Interdisciplinary Center of Theoretical Studies, Chinese Academy
of Sciences, P. O. Box 2735, Beijing 100080, China}
\affiliation{Center for Advanced study, Tsinghua University,
Beijing 100084, China}
\begin{abstract}

We present a general theory for the Fano resonance in Anderson
impurity systems. It is shown that the broadening of the impurity
level leads to an additional and important contribution to the
Fano resonance around the Fermi surface, especially in the mixed
valence regime. This contribution results from the
interference between the Kondo resonance and the broadened
impurity level. Being applied to the scanning tunnelling microscopic
experiments, we find that our theory gives a consistent and
quantitative account for the Fano resonance lineshapes for both Co
and Ti impurities on Au or Ag surfaces. The Ti systems are found
to be in the mixed valence regime.

\end{abstract}

\maketitle

The Fano resonance \cite{fano61} is a ubiquitous phenomenon
observed in different fields including atomic and condensed matter
physics. This resonance results from the interference between a
continuum and a discrete level embedded and is characterized by an
asymmetry factor $q$ for the lineshape \cite{fano61}. Recently,
the interest in the Fano resonance has been renewed in the study
of the Kondo effect by the scanning tunnelling microscope (STM)
measurements. Experimentally, the tunnelling spectra of 3d
transition metal atoms on noble metal surfaces manifest themselves
as Fano resonances near the Fermi level
$\varepsilon_F$\cite{madhavan98, li98, mano00, knorr02,
schneider02}. In the Kondo regime, for example in the systems of
Co atoms on Au \cite{madhavan98}, Cu \cite{mano00,knorr02}, or Ag
\cite{schneider02} surfaces, this resonance is believed to result
from the interference between the Kondo resonance and the
conduction electrons\cite{ujsaghy00, schiller00, plihal01,
cornaglia03, lin03}. However, in other impurity systems, such as
Ti atoms on Au\cite{jam00} or Ag \cite{nagaoka02} surfaces, the
lineshape appears much more complicated and cannot be explained
without invoking a broadened impurity level near the Fermi surface
\cite{jam00,nagaoka02}. In this case, the interference between the
Kondo resonance and the conduction electrons is dramatically
modified by the broadened impurity level and a microscopic picture
for the Fano resonance has not been established.

In this letter, we study the effect of the Fano resonance on the
density of states of conducting electrons in the Anderson impurity
systems by explicitly taking account of the broadening effect of
impurity levels. It is shown that the lineshape of Fano resonance
at the Fermi level is determined by two interference processes.
One is the interference between the Kondo resonance and the
broadened impurity level that serves effectively as an open
(quasi-continuum) channel, the other is the interference between
the Kondo resonance and the conduction band. While the
contribution from the former interference channel is very small in
the Kondo regime where the impurity levels lie well below or above
the Fermi energy, it becomes important in the mixed valence
regime, where the impurity levels are located within the linewidth
from the Fermi energy. In previous studies, attention has been
paid to the Fano resonance observed in the Kondo regime
\cite{madhavan98,mano00,knorr02,schneider02,li98}, but the recent
measurement data for Ti/Au\cite{jam00} and Ti/Ag \cite{nagaoka02}
appear not to fall into this category. We will show that by
incorporating the broadening effects, we can also give
quantitative account for the lineshapes of the Fano resonance in
these cases.

\begin{figure}[h]
\begin{center}
\includegraphics[width = 8.0cm, angle = 0]{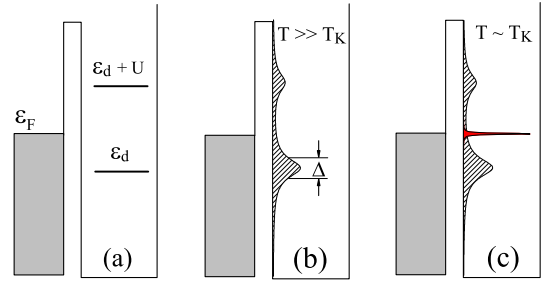}
\caption{Energy spectra for an Anderson impurity system: (a) The
conduction band with two impurity levels $\varepsilon_d$ and
$\varepsilon_d+U$ without hybridization; (b) With hybridization,
the two impurity levels are broadened with width $\Delta$; (c) In
the Kondo regime below the Kondo temperature $T_K$, a sharp Kondo
resonance is developed at the Fermi level.} \label{fig1}
\end{center}
\end{figure}

For an Anderson impurity system, as schematically shown in Fig.
\ref{fig1}, the hybridization of the impurity with the conduction
electrons leads to the broadening of the impurity levels. At low
temperatures, the conduction electrons screen the impurity spin,
and a Kondo resonance emerges near $\varepsilon_F$\cite{hewson}.
The impurity density of states is a superposition of the density
of states of the broadened levels and that of the Kondo resonance.
In the Kondo limit, the broadening effect can be neglected, and
the Fano resonance is predominantly due to the interference
between a Lorentzian-shaped Kondo resonance and the conduction
band. However, in the mixed valence regime, the density of states
at $\varepsilon_F$ due to the broadening becomes significant. This
broadened impurity level opens effectively an alternative
quasi-continuum channel and leads to an additional contribution to
the Fano resonance by interfering with the Kondo resonance.

To demonstrate the above picture, let us consider the Anderson
impurity model defined by \cite{anderson61}
\begin{eqnarray}
H &=& \sum_{k,\sigma}\varepsilon_k c^\dagger_{k\sigma} c_{k\sigma}
+ \sum_\sigma \varepsilon_{d}d^\dagger_\sigma d_\sigma + U
d^\dagger_{\uparrow}d_{\uparrow}d^\dagger_{\downarrow}
d_{\downarrow} \nonumber\\
&& + V\sum_{k,\sigma}(c^\dagger_{k\sigma} d_{\sigma} +
h.c.),\label{and}
\end{eqnarray}
where $c^\dagger_{k\sigma}$ and $d^\dagger_\sigma$ are the
creation operators for the conduction and impurity electrons,
respectively. $V$ is the hybridization integral and the impurity
level broadening is given by $\Delta = \pi \rho_{0} |V|^2$, where
$\rho_{0}$ is the density of states of conduction electrons at
$\varepsilon_F$.

The physical quantity measured by the STM is essentially the local
density of states of conduction electrons around the impurity site
($r = 0$). Due to the impurity scattering, the correction to the
retarded Green's function for the conduction electrons reads
\begin{equation}
\delta G_{c}(r,\omega) = |V|^2 G^0_{c}(r,\omega) G_{d}(\omega)
G^0_{c}(-r,\omega),
\end{equation}
where $G^0_{c}(r,\omega)$ and $G_{d}(\omega)$ are the retarded
Green's functions for the conduction electrons and impurity,
respectively. This leads to a correction to the local density of
states for conduction electrons:
\begin{eqnarray}
&&\delta \rho_c(r,\omega) = -\frac{1}{\pi} \mbox{Im}
\delta G_c(r,\omega)\label{gc} \nonumber \\
& = & -\Delta \rho_{0} \left[(q_c^2 - 1) \mbox{Im} G_d(\omega) - 2
q_c \mbox{Re} G_d(\omega)\right], \label{rhoc}
\end{eqnarray}
where $q_c = -\mbox{Re} G^0_c(r,\omega)/\mbox{Im}G^0_c(r,\omega)$.

In the Kondo limit, $G_d(\omega)$ consists of approximately three
well-separated Lorentzian poles \cite{ujsaghy00}. Two of them are
located at $\varepsilon_d$ and $\varepsilon_d + U$, and the third
one is the Kondo resonance at the Fermi level. In this case, Eq.
(\ref{gc}) can be recast into the standard form for the Fano
resonance with an asymmetry factor $q_c$\cite{ujsaghy00}. The Fano
resonance can then be interpreted as a result of the interference
between the Kondo resonance and the conduction band.

When $|\varepsilon_d -\varepsilon_F| $ is of order $\Delta$ or
smaller, the Lorentzian pole approximation for the Kondo resonance
is no longer valid since the broadening provides a new channel for
the interference and should be reflected in $G_d(\omega)$. Using
the Dyson equation, it can be shown that the impurity Green's
function $G_d(\omega)$ is given by
\begin{equation}
G_{d}(\omega) = G^0_{d}(\omega) + G^0_{d}(\omega) T_d(\omega)
G^0_{d}(\omega), \label{gd}
\end{equation}
where $T_d(\omega)$ denotes an effective scattering potential by
the Kondo resonance. $G^0_{d}(\omega)$ includes the contribution
from the hybridization:
\begin{equation}
G^0_{d}(\omega)= \frac{1-n/2}{\omega - \varepsilon_{d} + i\Delta}
+ \frac{n/2}{\omega - \varepsilon_{d} - U + i\Delta}, \label{gf0}
\end{equation}
where $n = \langle n_{d \uparrow}+ n_{d\downarrow} \rangle$ is the
average occupation number on the impurity site. From the imaginary
part of $G_d$, the density of states of the impurity is found to
be
\begin{eqnarray}
\rho_d (\omega) &=& \rho_{d,0} (\omega)  -\pi
\rho_{d,0}^2(\omega)\left[(q^2_{d} -
1) \, \mbox {Im}\,T_d(\omega)\right. \nonumber\\
&& \left. - 2 q_d \,\mbox{Re}\,T_{d}(\omega)\right], \label{gdim}
\end{eqnarray}
where $q_d = - \mbox{Re} G_d^0 (\omega) / \mbox{Im} G_d^0
(\omega)$ and $\rho_{d,0} (\omega) = -\mbox{Im} G_d^0(\omega
)/\pi$.

The scattering matrix $T_d$ is a complicated function of $\omega$.
However, around the Kondo energy, $T_d$ is mainly determined by
the Kondo resonance pole and is approximately given by
\begin{equation}
T_d(\omega) \approx \frac{\Gamma_K}{\pi \rho_{d,0}(\varepsilon_K
)} \frac{1}{\omega - \varepsilon_K + i \Gamma_K} + t_{incoh},
\label{tmatrix}
\end{equation}
where $\varepsilon_K$ is the energy of the Kondo resonance and
$\Gamma_K$ is its width. $t_{incoh}$ denotes the incoherent
contribution. Substituting (\ref{tmatrix}) into (\ref{gdim}) and
ignoring the $t_{incoh}$ term, we obtain
\begin{equation}
\rho_{d}(\omega) \approx \rho_{d,0}(\varepsilon_K ) \frac{(\tilde
\varepsilon + q_d)^2} {\tilde \varepsilon ^2 + 1} , \label{fanod}
\end{equation}
where $\tilde \varepsilon = (\omega - \varepsilon_K)/\Gamma_K$.
Eq. (\ref{fanod}) is a generalization of the standard formula for
the Fano resonance \cite{fano61}. The difference is that the
asymmetry factor $q_d$ is now $\omega$-dependent. It is natural to
interpret this formula as a result of the interference between the
Kondo resonance (serving as the discrete channel) and the
broadened impurity levels (serving effectively as the open
channel), although both channels belong to the same physical
object.

Equation (\ref{fanod}) captures the main feature of the density of
states of $d$-electrons in the Kondo limit, i.e., $|\varepsilon_d
- \varepsilon_K| \gg \Delta$. Around the Fermi level $|q_d|
\rightarrow \infty$ and $\rho_d$ takes a simple Lorentzian form
\begin{equation}
\rho_d(\omega ) \approx \frac{\Gamma_K}{\pi \Delta}
\frac{\Gamma_K}{(\omega -\varepsilon_K)^2 + \Gamma_K^2}.
\label{lorentz}
\end{equation}
In this case the broadening effect is weak. This is actually the
starting point of Ujsaghy {\it et al.} for the Fano resonance
analysis \cite{ujsaghy00}. On the other hand, if $|\varepsilon_d -
\varepsilon_K| \sim \Delta$, $\rho_d(\omega )$ is no longer a
Lorentzian around $\varepsilon_K$, neither can it be written as a
sum of a Lorentzian function like Eq. (\ref{lorentz}) and
$\rho_{d,0}$. This shows that the Fano resonance of conduction
electrons described by Eq. (\ref{gc}) is of a more complex form,
and cannot be expressed as a simple sum of two Fano resonances, as
assumed in Refs. [\onlinecite{jam00, nagaoka02}].

The above analysis indicates that two asymmetry factors, $q_c$ and
$q_d$, are needed in order to characterize the Fano resonance of
conduction electrons in the Anderson impurity model. The
introduction of this additional asymmetry factor $q_d$ is vital
for the description of the Fano resonance in the mixed valence
regime.

\begin{figure}[h]
\includegraphics[width = 8cm, angle = 0]{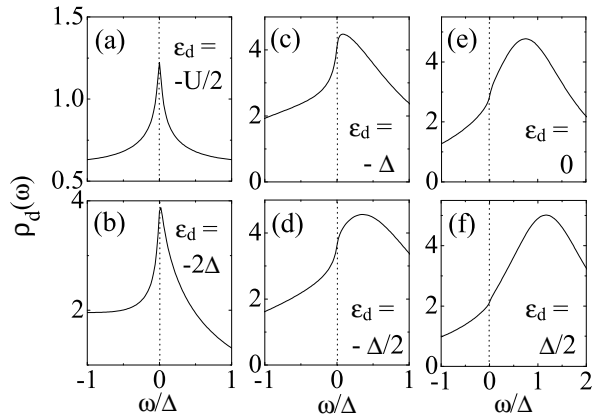}
\caption{$\rho_d(\omega)$ for different
$\varepsilon_d$ with $U=4\pi\Delta$. } \label{fig2}
\end{figure}
\begin{figure}[h]
\includegraphics[width = 8cm, angle = 0]{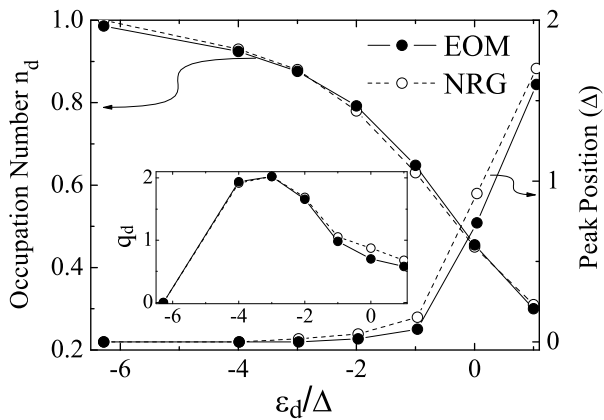}
\caption{Comparison of the results obtained with EOM and NRG
approaches for the impurity electron occupation number, the peak
position, and $q_d$ at the Fermi level (inset). The NRG results
are extracted from Table I and Figs. 5 and 8 in Ref.
\cite{costi94}.} \label{fig3}
\end{figure}

To elucidate more explicitly the broadening effect, we have
evaluated the impurity density of states using the equation of
motion (EOM) approach \cite{lacroix81, meir93}. Fig. \ref{fig2}
shows $\rho_d (\omega)$ for several limiting cases. In the Kondo
regime, the Kondo peak located at $\varepsilon_F$ is known to be
symmetric for the particle-hole symmetric case (Fig. \ref{fig2}a)
and asymmetric otherwise (Fig. \ref{fig2}b). However, in the mixed
valence or empty orbital regime (Fig. \ref{fig2}c-f), the sharp
Kondo resonance peak is washed out and is replaced by a kink at
$\varepsilon_F$ when $\varepsilon_d$ is above $\varepsilon_F$.
These are consistent with the numerical renormalization group
(NRG) results \cite{costi94}.
By further comparison with the NRG
results, we found that the height of the peak in $\rho_d$ obtained
with the EOM is underestimated. However, the results for the peak
position of $\rho_d$, the impurity electron occupancy, and
the key parameters for characterizing the Fano resonance $q_d$ are
all in good agreement with the NRG calculations (Fig. \ref{fig3}). In the
mixed valence regime, $\rho_d(\omega )$ cannot be written as a
Lorentzian, like Eq. (\ref{lorentz}), even after subtracting the
contribution from the broadened impurity levels.

Now let us apply our theory to the STM experiments for Co or Ti
atoms on noble metal surfaces. The tunnelling conductance measured
by STM is proportional to the local density of states of
conduction electrons $\rho_{0} +\delta \rho_c (\omega)$.

\begin{figure}[h]
\includegraphics[width = 8cm]{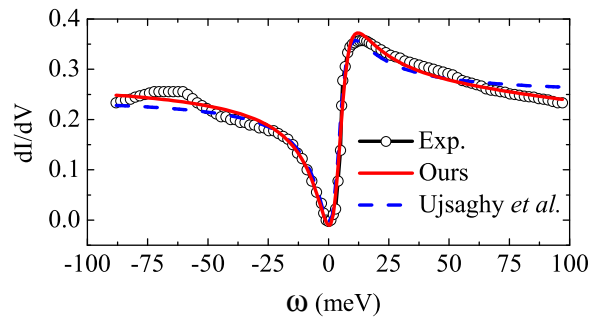}
\caption{Comparison between our theoretical result and the STM
measurement data for Co/Au \cite{madhavan98}. The result of Ref.
\cite{ujsaghy00} is also shown for comparison (dashed line).}
\label{fig4}
\end{figure}

For Co atoms on Au (111) surface \cite{madhavan98}, the model
parameters determined by density functional theory
\cite{ujsaghy00} are $\varepsilon_d -\varepsilon_F = -0.84\,eV$,
$U = 2.84\,eV$, $\Delta = 0.2\,eV$ and $n = 0.8$. This impurity
system is in the Kondo regime\cite{ujsaghy00}. However, as
$|\varepsilon_d -\varepsilon_F|$ is bigger than $\Delta$ only by a
factor of 4, the broadening effect can still have a small but
finite contribution. From the above parameters, the Fano factor
for the $d$-electrons is found to be $q_d (\varepsilon_F) \approx
2.6$. Taking all the above parameters as input, we have analyzed
the experimental data published in Ref. \cite{madhavan98} using
Eqs. (\ref{rhoc}-\ref{gf0}) and (\ref{tmatrix}). As shown in Fig.
\ref{fig4}, our result agrees well with the experimental data. The
fitting parameters used are $q_c = 1.4$, $\varepsilon_K = 4.0\,
meV$ and $\Gamma_K = 5.6\, meV$. Our result also agrees with the
theoretical calculation by Ujsaghy {\it et al.} \cite{ujsaghy00}.
However, the Fano factor $q_c$ we obtained is bigger than theirs
$q_c=0.66$, which indicates that the broadened impurity level has
a sizeable contribution to the Fano resonance, even if the system
is in the Kondo regime. Furthermore, the other parameters we
obtained are also slightly different from theirs. In addition, it
should be pointed out that although Eq. (3) in Ref.
\cite{ujsaghy00} captures the main feature of the Fano resonance
in the Kondo limit, the spectral weight of the $\varepsilon_d$
level, deduced from the parameters given in Ref. \cite{ujsaghy00},
is $Z_d \sim 11$ which is larger than the upper limit of $Z_d$
physically allowed.

For Ti/Au(111)\cite{jam00} and Ti/Ag(100)\cite{nagaoka02}, the
tunnelling spectra cannot be simply fitted by the standard Fano
formula. The authors of Refs. [\onlinecite{jam00, nagaoka02}]
proposed to use two Fano resonances to fit the tunnelling
conductance. In particular, they assumed that there is a narrow
Fano resonance, taking as the Kondo resonance, at $\varepsilon_F$,
and a broader Fano resonance slightly above $\varepsilon_F$,
originated from a bare Ti $d$-resonance. From the fitting, they
found that the normalized energy of Ti $d$-level
$\bar\varepsilon_d \sim 36\, meV$ and the broadening parameter
$\Delta = 127 \, meV$ for Ti/Au(111), and $\bar\varepsilon_d \sim
10\, meV$ and $\Delta = 78\, meV$ for Ti/Ag(100). Their results
indicate clearly that the broadening effect is rather strong in
these systems. However, as mentioned earlier, when the broadening
of the $d$-level is larger than the separation between
$\varepsilon_d$ and $\varepsilon_F$, it is not appropriate, even
approximately, to separate the bare $d$-resonance with the Kondo
resonance.

\begin{figure}[h]
\includegraphics[width = 8cm]{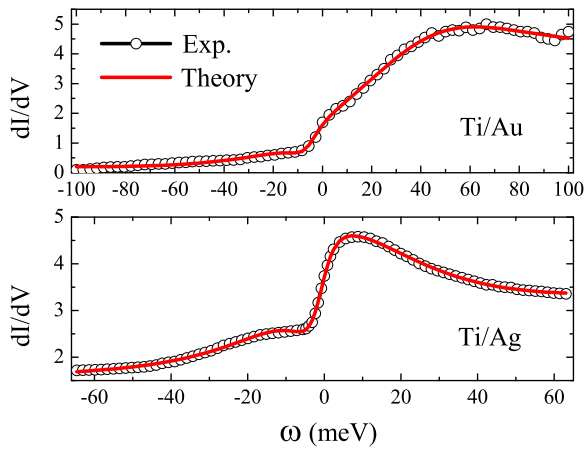}
\caption{Comparison between theoretical fitting curves and the STM
measurement data for Ti/Au(111) \cite{jam00} and Ti/Ag(100)
\cite{nagaoka02}.} \label{fig5}
\end{figure}

We have also analyzed the tunnelling spectra for Ti/Au(111) and
Ti/Ag(100) using our formula. Fig. \ref{fig5} compares our
theoretical results with the experimental data. Our theoretical
curves agree quantitatively with the experimental results
\cite{jam00,nagaoka02}. From the fitting (set $\varepsilon_F=0$),
we find that $n = 0.35$, $\varepsilon_d = 34.7\, meV$, $\Delta=
54.5\, meV$, $U = 81\, meV$, $\varepsilon_K = -5.8\, meV$,
$\Gamma_K = 7.4\, meV$, $q_c = 1.6$, and $q_d = -0.7$ for
Ti/Au(111), and $n = 0.6$, $\varepsilon_d = 0.97\, meV$, $\Delta=
29.2\, meV$, $U = 120.9\, meV$, $\varepsilon_K = -1.5\, meV$,
$\Gamma_K = 5.4\, meV$, $q_c = 0.9$, and $q_d = -0.13$ for
Ti/Ag(100). In agreement with \cite{jam00,nagaoka02}, we find that
$\varepsilon_d$ lies very closely to the Fermi level. Furthermore,
for both cases $|q_d| < q_c$, therefore the Fano resonance is
strongly modified by the broadening of the impurity levels. Our
result shows that Ti/Au and Ti/Ag are all in the mixed valence
regime. This can in principle be verified by density functional
band structure calculations \cite{step96}.

In summary, we have established a unified microscopic picture for
the Fano resonance in both Kondo and mixed valence regimes in the
Anderson impurity systems. It is shown that the broadened impurity
levels can effectively interfere with the Kondo resonance to
affect significantly the density of states measured by the STM,
especially in the mixed valence regime. Our theory gives a
quantitative account for the STM spectra of Co/Au as well as Ti/Au
or Ti/Ag systems. Our work is also of great interest for the
interpretation of experimental data in more complex impurity
systems explored in recent years, such as quantum dots
\cite{tk88,glaz88}.

We are grateful to M. F. Crommie for kindly providing us with the
experimental data. We also acknowledge fruitful discussions with
B. G. Liu, Y. L. Liu and G. M. Zhang and useful correspondence
with T. A. Costi, A. Zawadowski, and O. Ujsaghy. This work is
supported in part by the Special Funds for Major State Basic
Research Projects of China and by the National Natural Science
Foundation of China.


\begin{thebibliography}{99}
\bibitem{fano61} U. Fano, Phys. Rev. {\bf 124}, 1866 (1961).

\bibitem{madhavan98} V. Madhavan, W. Chen, T. Jamneala, M. F. Crommie, and N. S. Wingreen,
Science {\bf 280}, 567 (1998); Phys. Rev. B {\bf 64}, 165412 (2001).

\bibitem{li98} J. Li, W. D. Schneider, R. Berndt, and B. Delley,
Phys. Rev. Lett. {\bf 80}, 2893 (1998).

\bibitem{mano00} H. C. Manoharan, C. P. Lutz, and D. M. Eigler, Nature {\bf 403}, 512 (2000).

\bibitem{knorr02} N. Knorr, M. A. Schneider, L. Diekhoner, P. Wahl, and K. Kern, Phys. Rev. Lett. {\bf 88}, 096804 (2002).

\bibitem{schneider02} M. A. Schneider, L. Vitali, N. Knorr, and K. Kern, Phys. Rev. B {\bf 65}, 121406 (2002).

\bibitem{jam00} T. Jamneala, V. Madhavan, W. Chen, and M. F. Crommie, Phys. Rev. B {\bf 61}, 9990 (2000).

\bibitem{nagaoka02} K. Nagaoka, T. Jamneala, M. Grobis, and M. F. Crommie, Phys. Rev. Lett. {\bf 88}, 077205 (2002).

\bibitem{ujsaghy00} O. Ujsaghy, J. Kroha, L. Szunyogh, and A. Zawadowski, Phys. Rev. Lett. {\bf 85}, 2557 (2000).

\bibitem{schiller00} A. Schiller and S. Hershfield, Phys. Rev. B {\bf 61}, 9036 (2000).

\bibitem{plihal01} M. Plihal and J. W. Gadzuk, Phys. Rev. B {\bf 63}, 085404 (2001).

\bibitem{cornaglia03} P. S. Cornaglia and C. A. Balseiro, Phys. Rev. B {\bf 67}, 205420 (2003).

\bibitem{lin03} C. Y. Lin, A. H. Castro Neto, and B. A. Jones, cond-mat/0307185 (2003); J. Merino and O. Gunnarsson,
cond-mat/0308103 (2003).

\bibitem{anderson61}P. W. Anderson, Phys. Rev. {\bf 124}, 41 (1961).

\bibitem{hewson} A. C. Hewson, {\it The Kondo Problem to Heavy Fermions} (Cambridge University Press, Cambridge, U.K., 1993).

\bibitem{costi94} T. A. Costi, A. C. Hewson, and V. Zlatic, J. Phys.: Condens. Matter {\bf 6}, 2519 (1994).

\bibitem{lacroix81} C. Lacroix, J. Phys. F {\bf 11}, 2389 (1981);
H.-G. Luo, Z. -J. Ying, and S.-J. Wang, Phys. Rev. B {\bf 59}, 9710 (1999).

\bibitem{meir93} Y. Meir and N. S. Wingreen, and P. A. Lee, Phys. Rev. Lett. {\bf 70}, 2601 (1993); N. S. Wingreen and Y. Meir,
Phys. Rev. B {\bf 49}, 11040 (1994).

\bibitem{step96} V. S. Stepanyuk, W. Hergert, K. Wildberger, R. Zeller, and P. H. Dederichs,
Phys. Rev. B {\bf 53}, 2121 (1996).

\bibitem{tk88} T. K. Ng and P. A. Lee, Phys. Rev. Lett. {\bf 61}, 1768 (1988).

\bibitem{glaz88} L. I. Glazman and M. E. Raikh,
Pis'ma Zh. Eksp. Teor. Fiz. {\bf 47}, 378 (1988) [JETP Lett. {\bf 47}, 452 (1988)].
\end{thebibliography}
\end{document}